\documentstyle[12pt,psfig]{article}
\def \bg #1 {\begin{tabular}{{#1}}} \def \nd  {\end{tabular}}
\def \dfrac #1#2 {\displaystyle\frac{#1}{#2}}
\begin{document}
\thispagestyle{empty}

\vspace{8.0cm}
 \begin{center}
{\Large { Analytic Approach to Perturbative QCD and }}

\vspace{0.2cm}
{\Large { Renormalization Scheme Dependence} }\\

\vspace{0.7cm}
{\large { I.L. Solovtsov and D.V. Shirkov }}

\vspace{0.3cm}
{\it Bogoliubov Laboratory of Theoretical Physics,
          Joint Institute for Nuclear Research,\\
         Dubna, Moscow Region, 141980, Russia}

\vspace{0.6cm}
%\today
\end{center}

\vspace{2cm}
\noindent
\begin{abstract}
We further develop the approach recently used to construct an
analytic ghost-free model for the QCD running coupling based on the
requirement of the $Q^2$-analyticity and
apply it to the process of $e^+e^-$ annihilation into hadrons to study
the renormalization scheme dependence of the $R(s)$ cross-section
ratio. \par
    By transforming the relevant QCD corrections up to the three-loop
level into the ``analytized" form we show that the $R_{\rm AA}(s)$
expression thus obtained is remarkably stable (as compared to the
conventional perturbative approach) with respect to the renormalization
scheme dependence for the whole low-energy region.
\end{abstract}

\newpage
\setcounter{page}{1}
{\bf 1.} The analytic model for the QCD running coupling has been
devised in Ref.~\cite{jinr96} on the basis of the K\"all$\acute{{\rm
e}}$n--Lehmann analyticity (that is causality). This expression
$\bar{\alpha}_{\rm an}(Q)$, by construction, has no unphysical
singularities, like a ghost pole, in the complex $Q^2$-plane and its
perturbation expansion precisely coincides with the usual perturbation
one. The simplest version (discussed in detail in Ref.~\cite{prl97}) has
no extra parameters. Its infrared limit as $Q^2\to 0$ is
finite and independent of the QCD scale parameter $\Lambda$.
The numerical value $\bar{\alpha}_{\rm an}(0) \simeq 1.4$ turns out to
be remarkably stable with respect to higher loop corrections.

The analytic running coupling, by prescription, is defined via the
spectral integral
\begin{equation} \label{a-spectral}
a_{\rm an}(Q^2)\equiv\frac{\bar{\alpha}_{\rm an}(Q^2)}{\pi}\,=
\,\frac{1}{\pi} \int_0^\infty d\sigma \frac{\rho (\sigma)}
{\sigma+Q^2-{\rm i}\epsilon} \, ,
\end{equation}
with the density $\rho (\sigma)$ calculated
``perturbatively" as an imaginary part of the usual (RG-summed) running
coupling. This spectral representation provides us with correct
analytic properties of the running coupling $a_{\rm an}(Q^2)$: it is
an analytic function in the complex $Q^2$-plane with a cut along the
negative part of the real axis.

The one-loop expression for the analytic running coupling
thus defined can be presented explicitly
\begin{equation}  \label{a1-loop}
{a}_{\rm an}^{(1)}(Q^2)\,=\,\frac{1}{\beta_1} \left[\frac{1}
{\ln Q^2/\Lambda^2}\,+ \,\frac{\Lambda^2} {\Lambda^2-Q^2}\right]\,
\end{equation}
with $\beta_1=(11-2f/3)/4$
and $f$, the number of active quarks. The ghost pole ``is removed" by
the second term, the term that is nonanalytic in the coupling constant
$a_{\mu}=(\beta_1\ln\,\mu^2/\Lambda^2)^{-1}$,
i.e., has a nonperturbative nature.

   Two- and three-loop expressions for $\bar{\alpha}_{\rm an}(Q)$
discussed in our previous publications have a more complicated
structure and can explicitly be presented only after some
approximation\footnote{For the two-loop case one can use
expression~(\ref{a1-loop}) with the substitution \par
 $\quad \quad \quad
 Q^2/\Lambda^2\to\exp\left[\ln\,{Q^2}/{\Lambda^2}+\beta_2/(\beta_1)^2
\sqrt{\ln^2\,{Q^2}/{\Lambda^2}+4\pi^2}\right] $ -- see Ref.~\cite{prl97}.}.
Their analysis has revealed an important feature -- reasonable stability
of the analytic coupling behaviour in IR with respect to higher
loop corrections. A very small difference between two- and three-loop
curves (see Figure at Ref.~\cite{prl97}) gives, in general, hope of stability
with respect to the scheme dependence.

  In this letter, we concentrate on the problem of the renormalization
scheme (RS) dependence. Along with the discussion in Ref.~\cite{qcd97},
we are going to apply the idea of analytic approach (AA) not only to
QFT objects, like a running coupling, but directly to observables.
Here, we consider $R(s)$, the well known cross-section ratio for the
process $e^+e^- \to ~hadrons$. This example is physically interesting
for the RS stability issue (see, e.g., discussion in Ref.~\cite{racz}
and references therein). At the same time, it is instructive from the
theoretical point of view, as the time-like region is involved. \par

{\bf 2.}  The QCD correction $r(s)$ to the physical (and, therefore,
RS-invariant) quantity\footnote{Here and below all numerical
coefficients are given for three active quarks that is adequate to the
``low-$Q$ physics" we are interested in.} $R(s)$ is usually  written down
\begin{equation}  \label{r(s)}
R(s)\,=\,2\,[1+r(s)]\, ; \,~\, r(s)\,=\,\tilde{a}(s)\,\left[1\,+
\,r_1\tilde{a}(s)\,+\,r_2\tilde{a}^2(s)\right]\,   \end{equation}
in terms of $\tilde{a}(s)$, ``the running coupling in the time-like region"
which, by assumption, is defined as a mirror  image
$\tilde{a}(-Q^2) =a(Q^2)$.
This expansion is parallel to the analogous one for the
RS-invariant Adler function
\begin{eqnarray}
 \label{d(Q2)}
D(Q^2)&=&Q^2\frac{d\Pi(Q^2)}{dQ^2}=2\,[1+d(Q^2)]\,  , \nonumber \\
d(Q^2)&=&a(Q^2)\,\left[1\,+ \,d_1a(Q^2)\,+\,d_2a^2(Q^2)\right]\,  ,
\end{eqnarray}
where $\Pi(Q^2)$ is the vector current correlator
connected with $R(s)$ by the  relation
\begin{equation} \label{DviaR}
d(Q^2)\,=\,Q^2\int_0^\infty\frac{ds}{(s+Q^2)^2}\,r(s)\,  .
\end{equation}
Here, $r_1=d_1$ but the second (and higher) coefficient differs
$r_2\,-\, d_2\,= - (\beta_1\pi)^2/3\,=\,-16.65 $
by the so-called ``$\pi^2$-term" which is the ``trace" of an  analytic
continuation  from the space-like to the time-like region.
Taking into account that $d(Q^2)$ is an analytic
function in the complex $Q^2$-plane with a cut along the negative part
of the real axis, the analytic continuation procedure
can be described by the ``reverse relation"
\begin{equation} \label{reverse}
r(s)=\,-\, \frac{1}{2\pi {\rm i}}
\int^{s+{\rm i}\varepsilon}_{s-{\rm i}\varepsilon}
\frac{d\sigma}{\sigma} \, d(-\sigma) \, ,      \end{equation}
where the contour is in the region of analyticity of $d(Q^2)$.

However, this straightforward combination of analytic continuation
with the ``mirror" RG summation preserves unphysical singularities,
like a ghost pole in the one-loop term and unphysical cuts for higher
loops. Instead of it we shall apply the AA.
\medskip

 To compare the results obtained in various RS's, one uses the
'cancellation index criterion' proposed by P.~Raczka~\cite{racz95}.
According to it, a set of ``natural" RS's can be introduced which
includes schemes for which the degree of cancellation between
different terms in the second RS-invariant~\cite{stev}
\begin{equation}    \label{rho2}
\rho_2\,=\, r_2\,+\,c_2\,-\,r_1^2\,-\,c_1\,r_1\,=\,\tilde{\rho}(c_i,\,r_i)
\end{equation}
is ``not too large". The degree of cancellation is measured by
the 'cancellation index'
\begin{equation} \label{CI}
C_R\,=\,\frac{1}{|\rho_2|}\left( \,|r_2|\,+\,|c_2|\,+\,r_1^2\,+
\,c_1\,|r_1| \right)\,\equiv\, C(c_i,\,r_i)\, ,
\end{equation}
and the usual quantitative criterion is~\cite{racz95}
\begin{equation} \label{piotr}
C_R \leq 2 \, .   \end{equation}

Note here, in order
to compare various RS's, one can also use, instead of index $C_R$,
Eq.~(\ref{CI}), the ``Euclidean index" $C_E\,=\,C(c_i,\,d_i)$
expressed via the Adler function coefficients $d_k$ that are more
primary than $r_k$ as they emerge directly from the Feynman diagram
calculation.  This change $C_R \to C_E$ with appropriate
redefinition of the second RS--invariant $\rho_2 \to
\tilde{\rho}(c_i,\,d_i)$  (see Ref.~\cite{KS}) will influence the
classification of ``well-behaved schemes". We shall
not discuss here this point in detail.

{\bf 3.}  To implement the AA, we will base upon Eq.~(\ref{DviaR})
for QCD correction to the Adler function
and exploit the analytic property of $d(Q^2)$ expressed in the form
\begin{equation}  \label{d-spectral}  %d_{\rm AA}(Q^2)\,=
 d(Q^2)\,=\,\frac{1}{\pi}\,\int_0^\infty\,\frac{d\,
\sigma}{\sigma\,+\,Q^2} \,\varrho^{\rm eff}(\sigma)\, .
\end{equation}
It is essential that the Adler function is defined in the Euclidean region
where  the renormalization group (RG) method can be applied directly.

On the other hand, using the reverse relation for $r(s)$,
Eq.~(\ref{reverse}), we have (see Ref.~\cite{m-sol})
\begin{equation} \label{r-d}
r_{\rm AA}(s) \,=\,\frac{1}{\pi}\,\int_s^\infty\,
\frac{d\,\sigma}{\sigma} \,\varrho^{\rm eff}(\sigma)\, .
\end{equation}

  Here, we introduce an effective spectral density $\varrho^{\rm eff}
(\sigma)$ which corresponds to the discontinuity of $d(Q^2)$ and
can be presented as
\begin{equation}   \label{rho-eff}
\varrho^{\rm eff}(\sigma)\,=\,\rho_0(\sigma)\,+\,d_1\,
\varrho_1(\sigma)\, +\,d_2\,\varrho_2(\sigma)\, ,
\end{equation}
where the first term is just the density for the running
coupling  and $\varrho_k(\sigma)$ are related to higher powers
$\sim a^{k+1}(Q^2)$ in the expansion with the coefficients~\cite{GKL}
$d_1^{\overline{{\rm MS}}}=1.64$ and $d_2^{\overline{{\rm MS}}}=6.371$
(the latter corresponds to $r_2^{\overline{{\rm MS}}}=-10.284$).

The coefficients $r_k$ and $d_k$ in Eqs.~(\ref{rho-eff}) and
(\ref{d(Q2)}) as well as $c_2$ in the running coupling $a(Q^2)$
depend on the RS choice. In the $\beta$-function
\begin{equation} \label{3-loop-beta}
Q^2\,\frac{\partial\,a}{\partial\,Q^2}\,=\,-\,\beta_1\,a^2\,\left(1\,+   % 13
\,c_1a\,+\,c_2a^2\right)\, ; \,~\, a=a(Q^2)\, ,
\end{equation}
we have  $\beta_1=9/4=2.25, \, c_1=1.778$ and, e.g.,
 $c_2^{\overline{\rm MS}}= 4.471$. From Eq.\,(\ref{3-loop-beta}), taking
into account the first RS-invariant
$\rho_1=\beta_1\ln (Q^2/\Lambda^2_i)-d_1^i$
and using a scale parameter $\Lambda_{\overline{\rm MS}}$
in the ${\overline{\rm MS}}$ scheme as a reference one, one
finds the relation for the running coupling
\begin{equation}
\label{eq-to-a}
\beta_1\,\ln\,\frac{Q^2}{\Lambda^2_{\overline{\rm MS}}}\,=\,
d_1^{\overline{\rm MS}}\,-\,d_1^i\,+\,\Phi(a^i(Q^2),\,c_2^i)
\end{equation}
with\footnote{The explicit expression for $\Phi(a,c_2) $
can be found, for example, in Ref.~\cite{max}.}
\begin{equation}  \label{Phi}
\Phi(a,c_2)\,=\,\frac{1}{a}\,+\,c_1\ln\,\frac{\beta_1\,a}{(1\,+
\,c_1\,a)}
+\,c_2\,\int_0^a\,\frac{dx}{(1\,+\,c_1x)\,(1\,+\,c_1x\,+\,c_2x^2)}\,
\end{equation}
and $i$, the RS index.

To calculate the QCD correction $r(s)$ within the conventional
approach, one takes Eq.~(\ref{eq-to-a}) with $Q^2\,\to s$ to define
the ``running coupling'' $\tilde a(s)$ in the time-like domain and
substitute it into the expression for $r(s)$, Eq.~(\ref{r(s)}). As it
has been noted above, this procedure encounters difficulties with analytic
properties of the usual running coupling. Within the AA, we arrive at
the same quantity, $r_{\rm AA}(s)$, from Eq.~(\ref{r-d}). Here, the
effective spectral function $\varrho^{\rm eff}(\sigma)$ is defined as
an imaginary part of the expression (\ref{d(Q2)}), with the
function $a(Q^2)$ being replaced by the complex function $a(-\sigma)$
which has to be found numerically as a solution of Eq.~(\ref{eq-to-a})
with the replacement $\ln (Q^2/\Lambda_{\overline{ \rm MS} }^2) \to
\ln (\sigma/\Lambda_{\overline{ \rm MS }}^2) - {\rm i}\pi $.
\smallskip

To discuss the RS--dependence, we consider two examples of
``well-behaved" RS's: the scheme $A$  with the parameters
$r_1^{(A)}=-3.2 $ and $c_2^{(A)}=0$ (the so-called 't Hooft scheme)
and $\overline{\rm MS}$ scheme with
$r_1^{(\overline{\rm MS})}=1.64 \, ,\, c_2^{(\overline{\rm MS})}=4.47 \,$.
Both the schemes have a sufficiently small value of the cancellation
index $C_R \simeq 2$, which is close to that for the so-called
optimal RS based on the principle of minimal sensitivity~\cite{stev}
(see the discussion of this subject in Ref.~\cite{ell}).

%%%%%%%%%%  FIGURE - "RS-dep"  %%%%%%%%%%%%%%%%%%%%
	        \begin{figure}[pbt]
\centerline{ \psfig{file=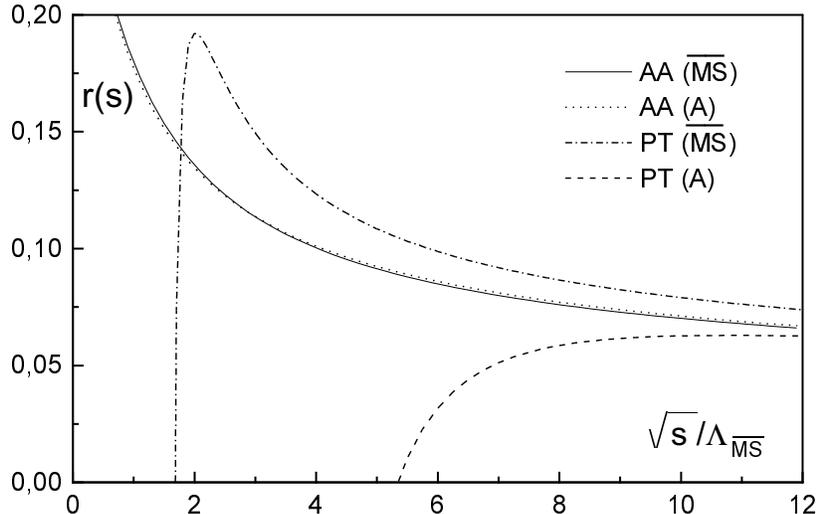,width=12.7cm}}
	 \caption{  {\sl
Plot of the QCD correction $r(s)$
calculated in the cases of perturbation theory ({\rm PT}) and the analytic
approach ({\rm AA}) in two different RS's with approximately the same
cancellation index $C_R\simeq 2$: A and $\overline{\rm MS}$.
			        }     }
	  \label{RS-dep}
          \end{figure}
%%%%%%%%%%%%%%%%%%%%%%%%%%%%%%%%%%%%%%%%%%%%%%%%%%%%%%%%%%%%%%
\medskip

In Fig.~\ref{RS-dep}, we plot the QCD correction $r(s)$ as a function
of $~\sqrt{s}/\Lambda_{\overline{\rm MS}}~$ for these two schemes
in usual treatment, as it was considered, e.g., in Refs.~\cite{racz,ckl,ms}
and within the AA. One can see that the analytically
improved result for $R(s)$ obeys a stable behaviour for the whole
interval of energies being  practically scheme independent.

Note also that the values of the parameter $\Lambda$ in the AA and in
perturbation theory are not equal to each other due to nonperturbative
contribution to the AA running coupling for the same value of the coupling
constant at some reference point. The value of $\Lambda_{\rm AA}$ is
larger than $\Lambda_{\rm PT}$ by a factor of about two~\cite{qcd97}.
Hence, our AA curves and the $\overline{\rm MS}$ one, being considered
as functions of $\sqrt{s}$, will be rather close to each other at
$s\geq \,2$ GeV.

{\bf 4.}
In our previous paper~\cite{prl97}, we have demonstrated that the
analytically improved running coupling being a  smooth function in
the IR turns out to be remarkably stable there with respect to higher
loop corrections.

In this note, we have applied the same idea of the analytical approach
to study the problem of theoretical ambiguity coming from the RS
dependence for a physical quantity, $R(s)$, the cross-section ratio of
the process $e^+e^-$ to hadrons. In the next-to-next-to-leading order
the Adler function with the correct analytic properties has been
constructed. This fact allowed us to find the corresponding
``analytized" expression for the QCD correction $r(s)$ to $R(s)$.
In contrast to the conventional perturbative expansion this
procedure does not encounter any problems and can be performed in a
self-consistent way. We have found that our AA reduces the RS
dependence drastically: $R_{\rm AA}(s)$ thus obtained turns out to be
practically RS independent for the whole energy interval.
\smallskip

\vspace{0.2cm}
The authors would like to thank Drs. Andrei Kataev  for valuable criticism
and Olga Solovtsova for useful comments.
Partial support of I.S. by RFBR grant 96-02-16126 and of D.Sh.
by INTAS 93-1180 and RFBR 96-15-96030 grants is gratefully acknowledged.

\end{document}